\documentclass[showpacs,showkeys,amssymb,preprintnumbers,nofootinbib,superscriptaddress,notitlepage,eqsecnum]{revtex4}

\usepackage{subeqn}
\usepackage{graphicx}

\begin{document}
\title{Fermion Quasi-normal modes of the Kerr Black-Hole}

\author{W.~A.~Carlson}
\email[Email: ]{warren.carlson@students.wits.ac.za}
\affiliation{National Institute for Theoretical Physics; School of Physics, University of the Witwatersrand, Wits 2050, South Africa}
\author{A.~S.~Cornell}
\email[Email: ]{alan.cornell@wits.ac.za}
\affiliation{National Institute for Theoretical Physics; School of Physics, University of the Witwatersrand, Wits 2050, South Africa}
\author{B.~Jordan}
\email[Email: ]{blake.jordan@students.wits.ac.za}
\affiliation{National Institute for Theoretical Physics; School of Physics, University of the Witwatersrand, Wits 2050, South Africa}

\begin{abstract}
In this paper we study the fermion quasi-normal modes of a $4$-dimensional rotating black-hole using the WKB(J) (to third and sixth order) and the AIM semi-analytic methods in the massless Dirac fermion sector. These semi-analytic approximations are computed in a pedagogical manner with comparisons made to the numerical values of the quasi-normal mode frequencies presented in the literature. It was found that The WKB(J) method and AIM show good agreement with direct numerical solutions for low values of the overtone number $n$ and angular quantum number $l$.
\end{abstract}

\preprint{WITS-CTP-86}
\date{15th January 2012}
\maketitle


\section{Introduction}\label{sec:1}

\par For a number of years black-hole (BH) quasi-normal modes (QNMs) have been of great interest to both gravitational theorists and gravitational wave experimentalists.  QNMs have been studied in the contexts of ringing of astrophysical BHs where QNMs are relevant to gravitational wave astronomy \cite{Kokkotas:1999bd, Thorne:1997cw, Stark:1985da}, loop gravity interpretations \cite{Hod:1998vk}, lower dimensional BHs \cite{Calmet:2010vp}, and in the AdS/CFT correspondence \cite{Maldacena:1997re} where BHs are thermal states in the field theory and their decays correspond to a decay of a perturbation of this state \cite{Horowitz:1999jd}. As such, knowledge of QNMs will provide an understanding of the stability of BHs against external perturbations; where more recently the interest in the Hawking radiation and QNMs of BHs has stemmed from the idea that mini-$TeV$ scale BHs might be created in particle accelerators such as the Large Hadron Collider (LHC) \cite{Antoniadis:1998ig, ArkaniHamed:1998rs, ArkaniHamed:1998nn, Randall:1999ee, Randall:1999vf}. QNM oscillations have been found in perturbation calculations of particles falling into Schwarzschild \cite{Davis:1971gg} and Kerr BHs \cite{Detweiler:1979xr} and in the collapse of a star to form a black hole \cite{Cunningham:1979px}.

\par Let us first recall that QNMs are the damped, resonant, non-radial perturbations of BHs which are characterized by a spectrum of discrete, complex frequencies, with real part determining the oscillation frequency, and an imaginary part that determines the rate at which each mode is damped as a result of an emission. For a given kind of physical perturbation the complex frequencies are uniquely determined by the mass and angular momentum of the BH, the angular harmonic indices $(l, m)$ of the deformation, and the degree of the harmonic of the mode $n$. As has been shown in many contexts, the fundamental equations describing the perturbations of BHs can be reduced to a single second-order ordinary differential equation similar to the one-dimensional Schr\"odinger equation for a particle encountering a potential barrier. However, the nature of the potential precludes an exact, closed-form solution in terms of known functions. That is, our equations for wave propagation reduce to
\begin{equation}
\frac{d^2 \psi}{dx^2} + \left(\omega^2 - V(x) \right) \psi = 0 \; , \label{eqn:101}
\end{equation}
where $x$ is a spatial co-ordinate. The BH event-horizon is mapped to $- \infty$ in the radial co-ordinate, and the potential $V(x)$ is a function of the spatial co-ordinate only. Note that we have proceeded in the usual way by assuming a time dependence, $\displaystyle{\Psi(t, x) = e^{i \omega t} \psi(x)}$.

\par This separation is not restrictive since once we have a solution for Eq.(\ref{eqn:101}), a general time dependent solution can be given as a continuous Fourier transform of such a solution, where the form of Eq.(\ref{eqn:101}) parallels a normal mode analysis. We shall now restrict ourselves to asymptotically flat space-times such that the potential $V(x)$ is positive and satisfies
\begin{equation}
V(x) \rightarrow 0\; , \quad x \rightarrow \pm \infty\; . \label{eqn:102}
\end{equation}
Note that there exist no normal mode expansions for such systems since such potentials do not allow bound states.  From the form of the potential in Eq.(\ref{eqn:102}) we have that near to the boundaries $\pm \infty$ the solutions behave as plane waves,
\begin{equation}
\psi(x) \sim e^{\pm i \omega x}, \quad x \rightarrow \pm \infty \; . \label{eqn:103}
\end{equation}
Here in-going at the horizon means entering into the BH, whilst the QNM frequencies and the associated wave-functions $\psi$ (which are the solutions to Eq.(\ref{eqn:101})) are the QNMs specified by the boundary conditions of Eq.(\ref{eqn:103}).

\par The study of BH perturbations was initiated by Regge and Wheeler \cite{Regge:1957td} and extended by Zerilli \cite{Zerilli:1971wd}. Following the work by Vishveshwara \cite{Vishveshwara:1970cc}, Chandrasekhar and Detweiler \cite{Chandrasekhar:1975zza} pioneered numerical methods for the study of QNMs.  More recently, semi-analytic methods have proven useful in the study of BH perturbations with the WKB(J) method of Schutz, Iyer and Will \cite{Schutz:1985km, Iyer:1986np}, and more recently the Asymptotic Iteration Method (AIM) of Ciftci, Hall and Saad \cite{Ciftci:2005xn}, which is a valuable tool for computing solutions to general second order ordinary differential equations. 

\par In this note we extend to sixth order in the semi-analytic approximation the works of Seidel and Iyer \cite{Seidel:1989bp}, Konoplya \cite{Konoplya:2011qq}, Iyer and Will \cite{Iyer:1986np}, and following the works of Refs. \cite{Cho:2009wf, Cho:2009cj}, we extend to the fermion sector the Kerr BH QNMs.  Here we shall compare the results of direct numerical, WKB(J) approximation and the AIM computation of QNMs of fermionic BH excitations in the Kerr space-time. Note that we shall assign $c = \hbar = G = 1$ and $M = 1$ for the calculated QNM frequencies in this work, where this note is organised as follows: In Sec.\ref{sec:2} we shall give a brief overview of the Newman-Penrose (NP) formalism \cite{Newman:1961qr} and its application to the QNMs of massless spin-$\frac{1}{2}$ fields. Sec.\ref{sec:3} and \ref{sec:4} contain reviews of the semi-analytic methods employed here, whilst Sec.\ref{sec:5} contains the numerical results of the QNMs for spin-$\frac{1}{2}$ field in a Kerr space-time, followed by a short discussion of these results.


\section{Newman Penrose Formalism}\label{sec:2}

\par In order to develop a master equation as discussed in the previous section, for a fermion, we shall introduce the NP formalism \cite{Newman:1961qr}. Let $(\mathcal{M}, g)$ be a pseudo-Riemannian manifold with metric $g$ of dimension $n$, and denote the tangent space at $p \in \mathcal{M}$  by $T_{p} \mathcal{M}$.  A natural basis for $T_{p} \mathcal{M}$ is the set of co-ordinate partial derivatives $\mathbf{e}_{\mu} \equiv \partial_{\mu}$, although we may choose a basis as we wish. Therefore, we choose one which is orthonormal with the signature of the metric $g$ on $\mathcal{M}$, that is
$$g(\mathbf{e}_{a}, \mathbf{e}_{b}) = \eta_{ab} \; , $$
where $\eta_{ab}$ is a block diagonal matrix in the NP formalism. A vector may be expressed as a linear combination of basis vectors such that
\begin{equation}
\mathbf{e}_{\mu} = e^{a}_{\mu} \mathbf{e}_{a} \; , \label{eqn:201}
\end{equation}
and the operators $e_{a}^{\mu}$ are $n \times n$ invertable matrices which satisfy,
$$e^{a}_{\mu} e^{\nu}_{a} = \delta^{\nu}_{\mu} \; ,$$
where Latin letters index tetrad co-ordinates and Greek letters index manifold co-ordinates. The operator $e^{a}_{\mu}$, now called a \emph{vielbein}, is a projection operator from the co-ordinate basis to the non-co-ordinate basis. The NP formalism is a special case for the \emph{vielbein}, or tetrad, formalism in which tensors defined on a space-time are projected on to a complete basis vector set at each point $p$.  This basis set is the set of complex null vectors $\mathbf{l}, \mathbf{n}, \mathbf{m}$ and $\mathbf{\bar{m}}$, of which $\mathbf{l}$ and $\mathbf{n}$ are real and $\mathbf{m}$ and $\mathbf{\bar{m}}$ are complex conjugate.  By the null condition of this basis vector set, it is clear that
$$\mathbf{l} \cdot \mathbf{l} = \mathbf{n} \cdot \mathbf{n} = \mathbf{m} \cdot \mathbf{m} = \mathbf{\bar{m}} \cdot \mathbf{\bar{m}} = 0 \; .$$
In addition, these basis vectors satisfy the orthogonality relations,
$$\mathbf{l} \cdot \mathbf{m} = \mathbf{l} \cdot \mathbf{\bar{m}} = \mathbf{n} \cdot \mathbf{m} = \mathbf{n} \cdot \mathbf{\bar{m}} = 0 \; ,$$
as well as the normalisation conditions $\mathbf{l} \cdot \mathbf{n} = 1$ and $	\mathbf{m} \cdot \mathbf{\bar{m}} = -1$. The corresponding fundamental matrix $\eta_{ab}$ is
$$\eta_{ab} = \eta^{ab} =
\left(\begin{array}{cccc}
0 & 1 & 0 & 0 \\
1 & 0 & 0 & 0 \\
0 & 0 & 0 & -1 \\
0 & 0 & -1 & 0
\end{array}\right) \; ,$$
to which there corresponds a basis, $\mathbf{e}_{1} = \mathbf{l}$, $\mathbf{e}_{2} = \mathbf{n}$, $\mathbf{e}_{3} = \mathbf{m}$ and $\mathbf{e}_{4} = \mathbf{\bar{m}}$.  When used to define directional derivatives we assign to each basis vector a special symbol, $\mathbf{e}_{1} = D$, $\mathbf{e}_{2} = D^{*}$, $\mathbf{e}_{3} = \delta$, and $\mathbf{e}_{4} = \delta^{*}$.

\par Having developed the NP formalism, we begin our analysis by considering the Dirac equation in the null tetrad basis,
\begin{equation}
\left( -i \gamma^{\mu} \nabla_{\mu} + m \right) \Psi = 0 \; , \label{eqn:202}
\end{equation}
where $\nabla_{\mu}$ is the covariant derivative taking the form
\begin{equation}
\nabla_{\mu} = \partial_{\mu} - \frac{i}{4} {{\omega_{\mu}}^{a}}_{b} \eta_{ca} \gamma^{cb} \; . \label{eqn:203}
\end{equation}
The spin connection ${{\omega_{\mu}}^{a}}_{b}$ satisfies the \emph{vielbein} ansatz,
\begin{equation}
\partial_{\mu} e^{a}_{\nu} + {{\omega_{\mu}}^{a}}_{b} e^{b}_{\nu} - \Gamma^{\sigma}_{\mu \nu} e^{a}_{\sigma} = 0 \; , \label{eqn:204}
\end{equation}
and $\gamma^{ab} = \gamma^{a} \gamma^{b} - \gamma^{b} \gamma^{a} $ with $\gamma^{a}$ being the $a^{th}$ Dirac matrix. Note that the spin connection ${{\omega_{\mu}}^{a}}_{b}$ may be rewritten in terms of the affine connection as
\begin{equation}
{{\omega_{\mu}}^{a}}_{b} = e^{a}_{\nu} e^{\lambda}_{b} \Gamma^{\nu}_{\mu \lambda} -  e^{\lambda}_{b} \partial_{\mu} e^{a}_{\lambda} \; . \label{eqn:205}
\end{equation}
The spin connection becomes the affine metric connection under the following co-ordinate transformation,
\begin{equation}
\Gamma^{\lambda}_{\mu \nu} = e^{\lambda}_{a} \left( \partial_{\mu} e^{a}_{\nu} + {{\omega_{\mu}}^{a}}_{b} e^{b}_{\nu} \right) \; . \label{eqn:206}
\end{equation}

\par We shall focus our attention on the massless limit of Eq.(\ref{eqn:202}),
\begin{equation}
\gamma^{\mu} \nabla_{\mu} \Psi = 0 \; , \label{eqn:207}
\end{equation}
rewriting $\Psi$ in terms of Weyl spinnors $\psi$ and $\tilde{\psi}$, that is,
\begin{equation}
\Psi = \left( \begin{array}{c}
\psi \\
\tilde{\psi}
\end{array} \right) \; , \label{eqn:208}
\end{equation}
and substituting Eq.(\ref{eqn:208}) into Eq.(\ref{eqn:207}) yields the dynamical Weyl spinor equations,
\begin{equation}
{\sigma^{\mu}}_{a \dot{b}} \nabla_{\mu} \psi^{a} = 0 \; , \quad \mathrm{and} \quad 
{\tilde{\sigma}^{\mu}}_{a \dot{b}} \tilde{\nabla}_{\mu} \tilde{\psi}^{\dot{b}} = 0 \; , \label{eqn:209}
\end{equation}
in the chiral representation. $\sigma^{a}$ and $\tilde{\sigma}^{\dot{a}}$ are Pauli matrices, and $\nabla_{\mu}$ and $\tilde{\nabla}{\mu}$ are the the covariant derivatives of $\psi^{a}$ and $\tilde{\psi}^{\dot{a}}$,respectively. The Pauli matrices take on new forms in the vielbein formalism for curved space-time, $\rho^{\mu} = e^{\mu}_{a} \sigma^{a}$, $\rho_{\mu} = e_{\mu}^{a} \sigma_{a}$, $\tilde{\rho}^{\mu} = e^{\mu}_{a} \tilde{\sigma}^{a}$ and $\tilde{\rho}_{\mu} = e_{\mu}^{a} \tilde{\sigma}_{a}$, where $e^{\mu}_{a}$ are the vielbein basis elements and $\sigma^{\mu}$ are Pauli matrices.  Explicitly we find
\begin{equation}
\begin{array}{lcr}
{\rho^{i}}_{a \dot{b}} =
\left( \begin{array}{cc}
l^{\mu} & m^{\mu} \\
\bar{m}^{\mu} & n^{\mu}
\end{array} \right)
\quad & \quad \text{and} \quad & \quad
{\rho_{i}}^{a \dot{b}} =
\left( \begin{array}{cc}
n_{\mu} & - m_{\mu} \\
- \bar{m}_{\mu} & l_{\mu}
\end{array} \right)
\end{array} \; . \label{eqn:210}
\end{equation}

\par Following the approach of Ref. \cite{Chandrasekhar:1985kt}, the covariant derivative of Eqs.(\ref{eqn:207}) and (\ref{eqn:209}) may be written as
\begin{equation}
{\rho^{i}}_{a \dot{b}} \nabla_{\mu} \psi^{a}  =  0 \; , \quad \mathrm{and} \quad 
{\tilde{\rho}^{i}}_{a \dot{b}} \tilde{\nabla}_{\mu} \tilde{\psi}^{\dot{b}} = 0 \; . \label{eqn:211}
\end{equation}

\par Consider now the $4$-dimensional Kerr background space-time with line element
\begin{eqnarray}
ds^{2}& = & - \left( 1 - \frac{2Mr}{\rho^{2}} \right) d t^{2} - \left( \frac{4Mra \sin^{2} {\theta}}{\rho^{2}} \right) d t d \theta \nonumber \\ 
&& \hspace{0.5cm} + \left( \frac{\rho^{2}}{\Delta} \right) d r^{2} + \left( \rho^{2} \right) d \theta^{2} + \left(r^{2} + a^{2} + \frac{2Mra \sin^{2} {\theta}}{\rho^{2}} \right) \sin^{2} {\theta} d \phi^{2} \; , \label{eqn:212}
\end{eqnarray}
where $\rho^{2} = r^{2} + a^{2} \cos^{2} {\theta}$, $\Delta = r^{2} + a^{2} - 2 Mr$ and $a = \frac{J}{M} \in [0; 1)$ is the rotation parameter of the BH. It is useful to define $\bar{\rho} = r + i a \cos{\theta}$ and $\bar{\rho}^{* } = r - i a \cos{\theta}$ such that $\rho^{2} = \bar{\rho} \bar{\rho}^{*}$. The basis of the NP formalism as defined on the Kerr geometry as
\begin{eqnarray}
l_{j} = \frac{1}{\Delta} \left( \Delta, - \rho^{2}, 0, - a \Delta \sin^{2} {\theta} \right) \; & , \; & n_{j} = \frac{1}{2 \rho^{2}} \left( \Delta, \rho^{2}, 0, - a \Delta \sin^{2} {\theta} \right) \; , \nonumber \\
m_{j} = \frac{1}{\sqrt{2} \bar{\rho}} \left( -i a \sin{\theta}, 0, - \rho^{2}, - i \left( r^{2} + a^{2} \right) \sin{\theta} \right) \; & , \; & \bar{m}_{j} = m_{j}^{*} \; , \nonumber
\end{eqnarray}
and
\begin{eqnarray}
l^{j} = \frac{1}{\Delta} \left( r^{2} + a^{2}, \Delta, 0, a \right) \; & , \; & n^{j} =  \frac{1}{2 \rho^{2}} \left( r^{2} + a^{2}, - \Delta, 0, a \right) \; , \nonumber \\
m^{j} = \frac{1}{\sqrt{2} \bar{\rho}} \left( i a \sin{\theta}, 0, 1, \frac{i}{\sin{\theta}} \right) \; & , \; & \bar{m}^{j} = {m^{j}}^{*} \; . \nonumber
\end{eqnarray}
We may now define the Ricci rotation coefficients (spin coefficients in that which follows) with respect to the curved space-time,
\begin{equation}
\begin{array}{ccc}
\kappa = \omega_{311} \; , & \quad \rho = \omega_{314} \; , & \quad \epsilon = \frac{1}{2} \left( \omega_{211} + \omega_{341} \right) \; , \\
\sigma = \omega_{313} \; , & \quad \mu = \omega_{243} \; , & \quad \gamma = \frac{1}{2} \left( \omega_{212} + \omega_{324} \right) \; , \\
\lambda = \omega_{244} \; , & \quad \tau = \omega_{312} \; , & \quad \alpha = \frac{1}{2} \left( \omega_{213} + \omega_{344} \right) \; , \\
\nu = \omega_{242} \; , & \quad \pi = \omega_{241} \; , & \quad \beta = \frac{1}{2} \left( \omega_{213} + \omega_{343} \right) \; .
\end{array} \label{eqn:213}
\end{equation}
Recall that the complex conjugate of any spin coefficient is obtained by replacing the numerical index $3$ with the value $4$ and vice versa \cite{Chandrasekhar:1985kt}. The resulting equations are
\begin{eqnarray}
\left(D + \epsilon - \tilde{\rho} \right) \psi^{0} + \left(\delta^{*} + \pi - \alpha \right) \psi^{1} & =& 0 \; , \nonumber \\
\left(D^{*} + \mu - \gamma \right) \psi^{1} + \left(\delta + \beta - \tau \right) \psi^{0} & = &0 \; ,\nonumber \\
\left(D + \epsilon^{*} - \tilde{\rho}^{*} \right) \tilde{\psi}^{\dot{0}} + \left(\delta + \pi^{*} - \alpha^{*} \right) \tilde{\psi}^{\dot{1}} & =& 0 \; ,\nonumber \\
\left(D^{*} + \mu^{*} - \gamma^{*} \right) \tilde{\psi}^{\dot{1}} + \left( \delta^{*} + \beta^{*} - \tau^{*} \right) \tilde{\psi}^{\dot{0}} & =& 0 \; . \nonumber 
\end{eqnarray}

\par It is clear that basis vectors are mapped to derivative operators when applied to the tangent space as tangent vectors to the wave-functions $\psi (t, r, \theta, \phi) = e^{i(\omega t + m \phi)} \psi (r, \theta)$.  Therefore, we write
\begin{eqnarray}
\mathbf{l} =  D = \mathcal{D}_{0} \; & , \; & \mathbf{n} =  D^{*} = - \frac{\Delta}{2 \rho^{2}} \mathcal{D}^{\dag}_{0} \; , \nonumber \\
\mathbf{m} =  \delta = \frac{1}{\sqrt{2} \bar{\rho}} \mathcal{L}^{\dag}_{0} \; & , \; & \mathbf{\bar{m}} =  \delta^{*} = \frac{1}{\sqrt{2} \bar{\rho}^{*}} \mathcal{L}_{0} \; , \nonumber
\end{eqnarray}
where
\begin{eqnarray}
\mathcal{D}_{n} = \partial_{r} + i \frac{K}{\Delta} + 2n \frac{r - M}{\Delta} \; & , \; & \mathcal{D}^{\dag}_{n} = \partial{r} - i \frac{K}{\Delta} + 2n \frac{r - M}{\Delta} \; , \nonumber \\
\mathcal{L}_{n} = \partial_{\theta} + Q + \frac{n}{ \tan{\theta}} - \frac{i n a \sin{\theta}}{\bar{\rho}} \; & , \; & \mathcal{L}^{\dag}_{n} = \partial_{\theta} - Q + \frac{n}{\tan{\theta}} + \frac{i n a \sin{\theta}}{\bar{\rho}^{*}} \; , \nonumber
\end{eqnarray}
with $K = \left( r^{2} + a^{2} \right) \omega + a m$, and $Q = a \omega \sin{\theta} + \displaystyle \frac{m}{ \sin{\theta}}$.

\par The spin coefficients are linear combinations of basis vectors in the NP formalism. We may therefore express the collection of spin connections in terms metric tensor elements. We shall choose the representation of the spin connections so as to simplify the algebra. A suitable choice for this assignment of elements is
\begin{eqnarray}
\tilde{\rho} = - \frac{1}{\bar{\rho}^{*}} \; , & \; \beta = \frac{1}{2 \sqrt{2} \bar{\rho}^{*} \tan{\theta}} \; , & \; \pi = \frac{ia \sin{\theta}}{\sqrt{2} (\bar{\rho}^{*})^{2}} \; , \nonumber \\
\tau = - \frac{ia\sin{\theta}}{\sqrt{2} \rho^{2}} \; , & \; \mu = - \frac{\Delta}{2 \bar{\rho^{*}} \rho^{2}} \; , & \;\gamma = \mu + \frac{r - M}{2 \rho^{2}} \; , \nonumber \\
& \; \alpha = \pi - \beta^{*} \; , & \nonumber
\end{eqnarray}
which results in,
$$\kappa = \sigma = \lambda = \nu = \epsilon = 0 \; .$$
Now let $f_{1} = \bar{\rho}^{*} \psi^{0}$, $f_{2} = \psi^{1}$, $g_{1} = \tilde{\psi}^{\dot{1}}$, $g_{2} = - \bar{\rho}\tilde{\psi}^{\dot{0}}$. Following these assignments, it can be shown \cite{Chandrasekhar:1985kt} that the Dirac equations reduce to
\begin{subequations}
\label{eqn:214}
\begin{eqnarray}
\mathcal{D}_{0} f_{1} + \frac{1}{\sqrt{2}} \mathcal{L}_{\frac{1}{2}} f_{2} &=& 0 \; , \\
\Delta \mathcal{D}^{\dag}_{\frac{1}{2}} f_{2} - \sqrt{2} \mathcal{L}^{\dag}_{\frac{1}{2}} f_{1} &=&  0 \; , \\\mathcal{D}_{0} g_{2} - \frac{1}{\sqrt{2}} \mathcal{L}^{\dag}_{\frac{1}{2}} g_{1} &=&  0 \; , \\
\Delta \mathcal{D}^{\dag}_{\frac{1}{2}} g_{1} + \sqrt{2} \mathcal{L}_{\frac{1}{2}} g_{2} &=&  0 \; . 
\end{eqnarray}
\end{subequations}
Using Eq.(\ref{eqn:214}), we separate the Dirac equation into radial and angular parts by making the ansatz:
\begin{subequations}
\label{eqn:215}
\begin{eqnarray}
f_{1} (r, \theta) &=& R_{- \frac{1}{2}} (r) S_{- \frac{1}{2}} \; , \\
f_{2} (r, \theta) &=& R_{\frac{1}{2}} (r) S_{\frac{1}{2}} \; ,\\
g_{1} (r, \theta) &=&  R_{\frac{1}{2}} (r) S_{- \frac{1}{2}} \; , \\			
g_{2} (r, \theta) &=&  R_{- \frac{1}{2}} (r) S_{\frac{1}{2}} \; .
\end{eqnarray}
\end{subequations}
Substituting Eq.(\ref{eqn:215}) into Eq.(\ref{eqn:214}) yields,
\begin{subequations}
\label{eqn:216}
\begin{eqnarray}
\mathcal{L}_{\frac{1}{2}} S_{\frac{1}{2}} &=&  - \lambda S_{- \frac{1}{2}} \; , \\
\mathcal{L^{\dag}}_{\frac{1}{2}} S_{- \frac{1}{2}} &=&  \lambda S_{\frac{1}{2}} \; ,
\end{eqnarray}
\end{subequations}
and
\begin{subequations}
\label{eqn:217}
\begin{eqnarray}
\sqrt{\Delta} \mathcal D_{0} R'_{- \frac{1}{2}} &=& \lambda \sqrt{\Delta} R_{\frac{1}{2}} \; , \\
\sqrt{\Delta} \mathcal D^{\dag}_{0} \sqrt{\Delta} R_{- \frac{1}{2}} &=& \lambda R'_{- \frac{1}{2}} \; ,
\end{eqnarray}
\end{subequations}
where $R'_{ - \frac{1}{2}} = \sqrt{2} R_{\frac{1}{2}}$.

\par Decoupling Eq.(\ref{eqn:216}) yields,
\begin{equation}
\left[ \mathcal{L}_{ \pm \frac{1}{2}} \mathcal{L}_{ \pm \frac{1}{2}} + \lambda^{2} \right] S_{ \mp \frac{1}{2}} = 0 \; , \label{eqn:218}
\end{equation}
where $S_{\frac{1}{2}}$ is the \emph{adjoint} of $S_{- \frac{1}{2}}$ under the replacement of $\theta$ by $\pi - \theta$. Similarly, decoupling Eq.(\ref{eqn:217}) yields,
\begin{equation}
\left[ \Delta \mathcal{D}^{\dag}_{\frac{1}{2}} \mathcal{D}_{0} \right] R_{- \frac{1}{2}} = 0 \; , \label{eqn:219}
\end{equation}
and $\sqrt{\Delta} R_{ \frac{1}{2}}$ satisfies the complex conjugate relation. Once decoupled, Eqs.(\ref{eqn:218}) and (\ref{eqn:219}) are eigenvalue equations with known eigenvalues $\lambda$ and unknown eigen-functions $R_{ \pm \frac{1}{2}}$ and $S_{ \pm \frac{1}{2}}$.

\par For the purpose of computing QNMs we shall focus our interest on the radial equation for the spin-$\frac{1}{2}$ field. We map Eq.(\ref{eqn:219}) to the \emph{tortoise} co-ordinate $x$,
\begin{equation}
\left(\frac{dr}{dx}\right) = \frac{\Delta}{\bar{K}} \; , \label{eqn:220}
\end{equation}
where $\bar{K} = \displaystyle \frac{K}{\omega}$ eliminates the co-ordinate singularity at the horizon. After this co-ordinate transformation, operators $\mathcal{D}_{0}$ and $\mathcal{D}^{\dag}_{0}$ simplify to,
\begin{subequations}
\label{eqn:221}
\begin{eqnarray}
\mathcal{D}_{0} &=& \frac{\bar{K}}{\Delta} \left( \frac{d}{ d x} + i \omega \right) \; , \\
\mathrm{and} \mathcal{D}^{\dag}_{0} &=& \frac{\bar{K}}{\Delta} \left( \frac{d}{d x} - i \omega \right) \; .
\end{eqnarray}
\end{subequations}
The assignments $P_{\frac{1}{2}} = \sqrt{\Delta} R_{\frac{1}{2}}$ and $P_{-\frac{1}{2}} = R_{-\frac{1}{2}}$, in conjunction with Eq.(\ref{eqn:221}), lead to the new form of the radial equations
\begin{subequations}
\begin{eqnarray}
\left( \frac{d}{ d x} - i \omega \right) P_{\frac{1}{2}} &=& \lambda \frac{\sqrt{\Delta}}{\bar{K}} P_{- \frac{1}{2}} \; , \label{eqn:222a} \\
\left( \frac{d}{ d x} + i \omega \right) P_{- \frac{1}{2}} &=& \lambda \frac{\sqrt{\Delta}}{\bar{K}} P_{\frac{1}{2}} \; . \label{eqn:222b}
\end{eqnarray}
\end{subequations}
If we define,
\begin{equation}
Z_{\pm} = P_{\frac{1}{2}} \pm P_{- \frac{1}{2}} \; , \label{eqn:223}
\end{equation}
and combine Eqs.(\ref{eqn:222a}) and (\ref{eqn:222b}) we obtain
\begin{eqnarray}
\left( \frac{d}{d x} - \lambda \frac{\sqrt{\Delta}}{\bar{K}} \right) Z_{+} &=& i \omega Z_{-} \; ,\nonumber \\
\left( \frac{d}{d x} + \lambda \frac{\sqrt{\Delta}}{\bar{K}} \right) Z_{-} &=& i \omega Z_{+} \; , \nonumber
\end{eqnarray}
which may be separated to yield
\begin{equation}
\left( \frac{d^2 Z_{\pm}}{dx^2}\right) \left( \omega^{2} - V_{\pm}(x) \right) Z_{\pm} = 0\; , \label{eqn:224} 
\end{equation}
with
\begin{equation}
V_{\pm}(x) = \lambda^{2} \frac{\Delta}{\bar{K}^{2}} \pm \lambda \frac{d}{d x} \left( \frac{\sqrt{\Delta}}{\bar{K}} \right) \; . \label{eqn:225}
\end{equation}
Eq.(\ref{eqn:224}) is now in the form of Eq.(\ref{eqn:101}) with a potential given by Eq.(\ref{eqn:225}).  Eq.(\ref{eqn:224}) is the QNM master equation for a spin-$\frac{1}{2}$ field in a Kerr space-time. In the following two sections we shall review two semi-analytic approaches to solving these equations for an ultimate comparison with the direct numerical results of Refs. \cite{Jing:2005dt, Jing:2005pk}.


\section{WKB(J) Method}\label{sec:3}

\par
\begin{figure}[h]
\centering
\includegraphics[width=7cm]{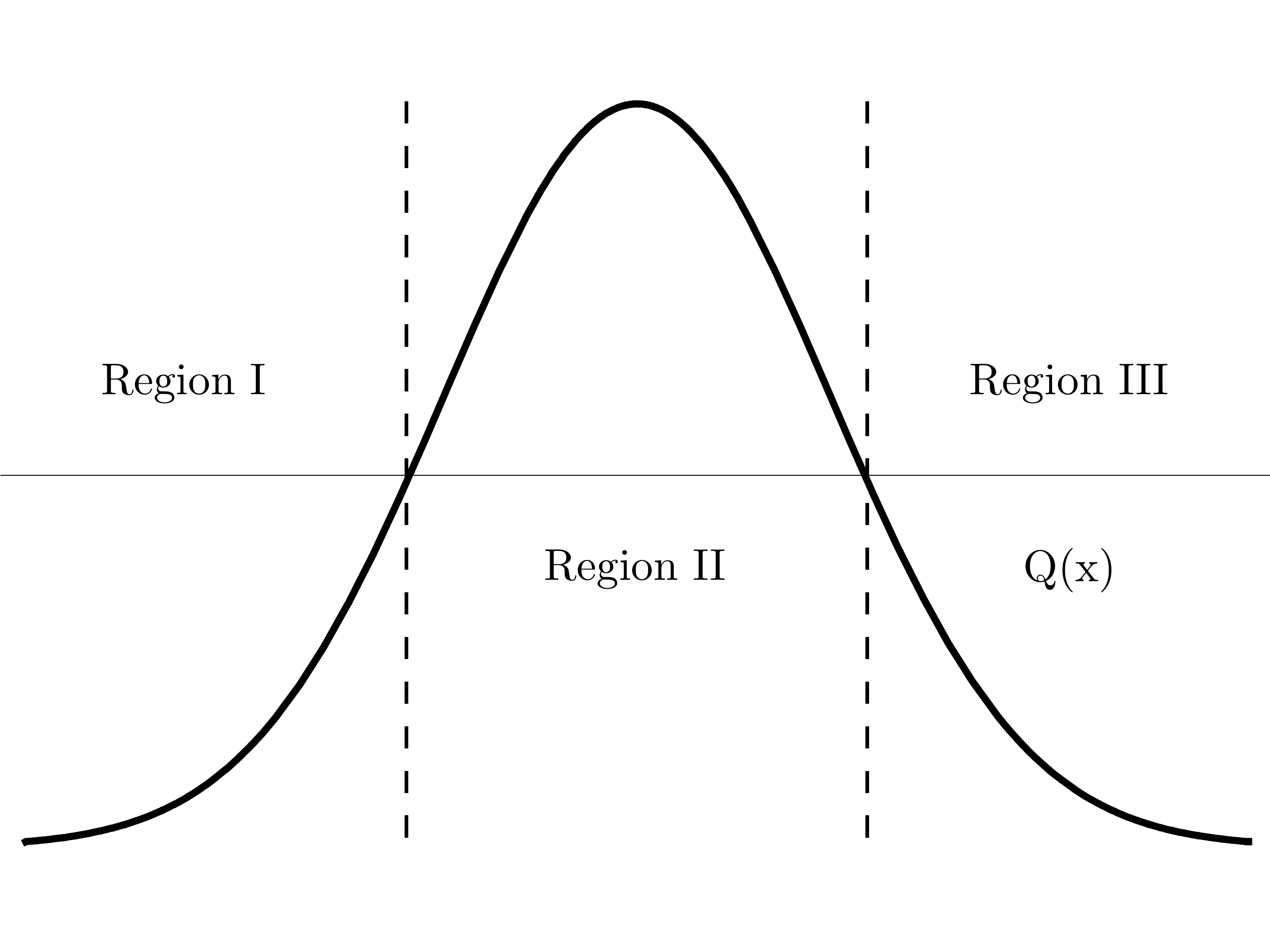}
\caption{\it The reduced potential barrier as used in the WKB(J) method.}\label{fig:1}
\end{figure}

 We shall begin with the WKB(J) approximation of Schutz, Iyer and Will \cite{Schutz:1985km, Iyer:1986np} which is a general method for generating approximate solutions to ordinary linear second order differential equations and has been used extensively in various BH cases (see Ref. \cite{Konoplya:2011qq} and references therein). For further discussion of this method, see Refs. \cite{Iyer:1986np, Seidel:1989bp, Konoplya:2011qq, Konoplya:2003ii}. We may use the WKB(J) method to generate approximate solutions to differential equations of the form
\begin{equation}
\left( \frac{d^2 \Psi}{dx^2} \right) + Q(x) \Psi = 0 \; , \label{eqn:301}
\end{equation}
for a smooth function, at $\mathcal{O}(1)$, $Q(x) = \omega^{2} - f(x)$ with $\omega \in \mathbb{C}$. Now assume that $\Psi$ has a mode expansion of the form,
\begin{equation}
\Psi(x, \epsilon) = A e^{ \sum_{n = 0}^{\infty} { i S_{n}(x) \epsilon^{n - 1}}} \; , \label{eqn:302}
\end{equation}
where $A \in \mathbb{C}$ and $\epsilon \ll 1$ tracks the order of the expansion. By substituting Eq.(\ref{eqn:302}) into Eq.(\ref{eqn:301}), and equating powers of $\epsilon$, yields
\begin{equation}
S_{0} (x) = \pm i \int^{x} {d t \ \sqrt{Q(t)}} \; , \label{eqn:303} 
\end{equation}
and
\begin{equation}
S_{1} (x) = - \frac{1}{4} \ln{Q(x)} \; , \label{eqn:304}
\end{equation}
where the two choices in sign in Eq.(\ref{eqn:303}) correspond to either incoming or outgoing waves at $\pm \infty$. When $x \rightarrow + \infty$, $Q(x) \rightarrow \omega^{2}$ such that $S_{0} \rightarrow + i \omega x$ for the outgoing wave to infinity and $S_{0} \rightarrow - i \omega x$ for the incoming wave from infinity. Similarly, for $x \rightarrow - \infty$, $S_{0} \rightarrow + i \omega x$ for the wave incoming from $x \rightarrow - \infty$, while $S_{0} \rightarrow - \omega x$ to a wave outgoing to $x \rightarrow - \infty$, see Fig.\ref{fig:1}. Designating these four solutions,
\begin{equation}
\begin{array}{ccc}
\Psi^{I}_{+} \sim e^{+ \omega x} \; , & \Psi^{I}_{-} \sim e^{- \omega x} \; , & x \rightarrow + \infty \\
\Psi^{III}_{-} \sim e^{+ \omega x} \; , & \Psi^{III}_{-} \sim e^{- \omega x} \; , & x \rightarrow - \infty
\end{array} \label{eqn:305}
\end{equation}

\noindent The general solutions to regions I and III are
\begin{equation}
\begin{array}{ccl}
\Psi \sim Z_{I}^{IN} \Psi_{I}^{-} + Z_{I}^{OUT} \Psi^{I}_{+} \; , & \text{region~I} \; , \\
\Psi \sim Z_{III}^{IN} \Psi_{II}^{-} + Z_{III}^{OUT} \Psi^{I}_{+} \; , & \text{region~III} \; , 
\end{array} \label{eqn:306}
\end{equation}
where $Z_{I}^{-}, Z_{I}^{+}, Z_{III}^{-}, Z_{III}^{+} \in \mathbb{C}$.

\par The amplitudes at $+ \infty$ are related to amplitudes at $- \infty$ through the linear relation
\begin{equation}
\left( \begin{array}{c}
Z_{III}^{+} \\
Z_{III}^{-}
\end{array} \right) =
\left(\begin{array}{cc}
S_{11} & S_{12} \\
S_{21} & S_{22}
\end{array}\right) \cdot
\left( \begin{array}{c}
Z^{I}_{+} \\
Z^{I}_{-}
\end{array} \right) \; . \label{eqn:307}
\end{equation}
We match the solutions in regions I and III with a solution in region II through the two turning points of $Q(x) = 0$. If the turning points are closely spaced ($- Q(x)_{max} \ll Q(\pm \infty)$), then the solution in region II can be well approximated by a Taylor series expansion,
\begin{equation}
Q(x) = Q(x_{0}) + \frac{1}{2} Q''(x_{0}) \left(x - x_{0} \right)^{2} + \mathcal{O} \left( \left(x - x_{0} \right)^{3} \right) \; , \label{eqn:308}
\end{equation}
where $x_{0}$ is the location of the maximum of the function $Q(x)$ and the prime notation signifies differentiation with respect to $x$. We shall assign $Q_{0} = Q \left( x_{0} \right)$ in what follows.  Region II corresponds to
\begin{equation}
\left| x - x_{0} \right| < \sqrt{ \frac{-2 Q_{0}}{Q_{0}''} } \approx \epsilon^{\frac{1}{2}} \; . \label{eqn:309}
\end{equation}
We introduce new functions
\begin{eqnarray}
k &= & \frac{1}{2} Q_{0}''  \; , \nonumber \\
t &= & (4 k)^{\frac{1}4{}} e^{\frac{i \pi}{4}} \left(x - x_{0} \right) \; , \label{eqn:310} \\
\mathrm{and} \; \; \nu + \frac{1}{2} &=& \frac{- i Q_{0}}{\sqrt{2 Q_{0}'' }} \; . \label{eqn:311}
\end{eqnarray}
Rewriting Eq.(\ref{eqn:301}) as
\begin{equation}
\left( \frac{d^2\Psi}{dt^2}\right) + \left( \nu + \frac{1}{2} - \frac{1}{4} t^{2} \right) \Psi = 0 \; , \label{eqn:312}
\end{equation}
which has, is terms of parabolic cylinder functions $F_{\nu} (t)$, the general solution
\begin{equation}
\Psi (t) = A F_{\nu} + B F_{\nu - 1} (it) \; . \label{eqn:313}
\end{equation}
It can be shown that the large $|t|$ asymptotics of this solution give an $S$-matrix,
\begin{equation}
\left( \begin{array}{c}
Z_{III}^{+} \\
Z_{III}^{-}
\end{array} \right) =
\left( \begin{array}{cc}
e^{i \pi \nu} & \frac{i R^{2} e^{i \pi \nu } (2 \pi)^{\frac{1}{2}}}{\Gamma (\nu + 1)} \\
\frac{R^{-2} (2 \pi)^{\frac{1}{2}} }{\Gamma (- \nu)} & - e^{i \pi \nu}
\end{array}\right)\cdot
\left( \begin{array}{c}
Z^{I}_{+} \\
Z^{I}_{-}
\end{array} \right) \; , \label{eqn:314}
\end{equation}
where $\Gamma$ is the Gamma function and $R$ is a general exponential function depending on the nature of the potential $Q$. When expanding to higher orders, the $S$ matrix has the same general form as shown in Eq.(\ref{eqn:314}), with modified expressions for the off-diagonal elements. Note that for a BH no waves are reflected by the horizon, such that $Z^{III}_{-} = 0$, and due to the QNM boundary conditions there are no waves incoming from infinity, $Z^{I}_{-} = 0$. These conditions are satisfied in Eq.(\ref{eqn:307}) only if $\Gamma(- \nu) = 0$. Consequently, $\nu \in \mathbb{Z}$, and
\begin{equation}
\alpha = \nu + \frac{1}{2} = \frac{- i Q_{0}}{\sqrt{2 Q_{0}''}} \; , \quad n = 0, 1, 3, \dots \; . \label{eqn:315}
\end{equation}
The above equation gives the quantisation condition which labels QNMs by an overtone number $n = - \nu$. Since the coefficients in the $S$-matrix that connect amplitudes near the horizon with those at infinity depend only on  $\alpha$ (related to the overtone number $n$), we need only expand Eq.(\ref{eqn:302}) to higher orders in $\epsilon$ to obtain the WKB formula to higher orders. The result has the form
\begin{equation}
\frac{i Q_{0}}{\sqrt{Q_{0}'' }} - \Lambda_{2} - \Lambda_{3} - \Lambda_{4} - \Lambda_{5} - \Lambda_{6} = n + \frac{1}{2} \; , \label{eqn:316}
\end{equation}
where the correction terms $\Lambda_{4}, \Lambda_{5}, \Lambda_{6}$ can be found in Ref. \cite{Konoplya:2003ii} and the included URL.


\section{Asymptotic Iteration Method}\label{sec:4}

\par Our second semi-analytic technique is the AIM, which produces an exact solution to an ordinary second order partial differential equation under the assumption that the coefficients in the Taylor expansion of the differential equation form an asymptotic series. To see this, consider the homogeneous linear second-order differential equation for the function $\chi (x)$,
\begin{equation}
\chi'' = \lambda_{0} (x) \chi' + s_{0} (x) \chi, \; , \label{eqn:401}
\end{equation}
where $\lambda_{0} (x)$ and $s_{0} (x)$ are functions in $C_{\infty} (a, b)$. In order to find a general solution to this equation we rely on the symmetric structure of the right-hand side of Eq.(\ref{eqn:401}) \cite{Ciftci:2005xn}. If we differentiate Eq.(\ref{eqn:401}) with respect to $x$, we find that
$$\chi''' = \lambda_{1} (x) \chi' + s_{1} (x) \chi \; ,$$
where $\lambda_{1} = \lambda'_{0} + s_{0} + ( \lambda_{0})^{2}$ and $s_{1} = s'_{0} + s_{0} \lambda_{0}$. Taking the second derivative of Eq.(\ref{eqn:401}) yields
$$\chi'''' = \lambda_{2} (x) \chi' + s_{2} (x) \chi \; ,$$
where $\lambda_{2} = \lambda'_{1} + s_{1} + \lambda_{0} \lambda_{1}$, and $s_{1} = s'_{0} + s_{0} \lambda_{0}$. The $(n + 1)^{th}$ and the $(n + 2)^{th}$ derivatives, $n = 1, 2, \dots$, are found iteratively,
\begin{equation}
\chi^{ (n + 1) } = \lambda_{n - 1} (x) \chi' + s_{n - 1} (x) \chi \; . \label{eqn:402}
\end{equation}
The crucial observation in the AIM is that differentiating the above equation $n$ times with respect to $x$, leaves a symmetric form for the right hand side:
\begin{equation}
\chi^{(n+2)} = \lambda_{n}(x) \chi' + s_{n}(x) \chi \; , \label{eqn:403}
\end{equation}
where,
\begin{subequations}
\label{eqn:404}
\begin{equation}
\lambda_{n}(x) = \lambda'_{n - 1} (x)+ s_{n - 1}(x) + \lambda_{0} (x) \lambda_{n-1} (x) \; , 
\end{equation}
and
\begin{equation}
s_{n} (x) = {s'}_{n - 1} (x) + s_{0} (x) \lambda_{n - 1} (x) \; .
\end{equation}
\end{subequations}
For sufficiently large $n$
\begin{equation}
\frac{s_{n} (x)}{\lambda_{n} (x)} = \frac{s_{n-1} (x)}{\lambda_{n-1} (x)} \equiv \beta(x) \; , \label{eqn:405}
\end{equation}
where from the ratio of the $(n + 1)^{th}$ and the $(n + 2)^{th}$ derivatives, we have
\begin{equation}
\frac{ d }{ d x } \ln{ \chi^{(n + 1)} } = \frac{ \chi^{(n + 2)} }{ \chi^{(n + 1)} } = \frac{ \lambda_{n} \left( \chi' + \frac{s_{n}}{\lambda_{n}} \chi \right) }{ \lambda_{n - 1} \left( \chi' + \frac{s_{n - 1}}{\lambda_{n - 1}} \chi \right)} = \frac{ d }{d x} \ln{ \chi^{(n + 1)} } = \frac{ \lambda_{n} }{ \lambda_{n - 1} } \; . \label{eqn:406}
\end{equation}
This then yields,
\begin{equation}
\chi^{(n+1)}(x) = C_{1} \exp{ \left( \int^{x} d x' \frac{ \lambda_{n} (x')}{ \lambda_{n - 1} (x')} \right)} = C_{1} \lambda_{n - 1} \exp{ \left( \int^{x} d x' (\beta + \lambda_{0}) \right)} \; , \label{eqn:407}
\end{equation}
where $C_{1}$ is the integration constant and the right-hand side of Eq.(\ref{eqn:404}) and the definition of $\beta(x)$ have been used. Substituting this into Eq.(\ref{eqn:402}), we obtain the first-order differential equation
\begin{equation}
\chi' + \beta \chi = C_{1} \exp{ \left( \int^{x} d x' (\beta + \lambda_{0}) \right)} \; . \label{eqn:408}
\end{equation}
The general solution can be determined as \cite{Ciftci:2005xn}
\begin{equation}
\chi(x) = \exp{ \left[ - \int^{x} d x' \beta (x') \right] \left( C_{2} + C_{1} \int^{x} d x' \exp{ \int^{x'} d x''\left[ \lambda_{0} (x'')+ 2 \beta (x'') \right] } \right)} \; . \label{eqn:409}
\end{equation}
The integration constants, $C_{1}$ and $C_{2}$, can be determined by an appropriate choice of normalisation. Note that for the generation of exact solutions $C_{1} = 0$. Eq.(\ref{eqn:405}) gives the sequence termination condition,
\begin{equation}
s_{n} \lambda_{n - 1} - s_{n - 1} \lambda_{n} = 0 \; ,\nonumber
\end{equation}
which is also the QNM ``quantisation condition'' as given in Ref. \cite{Barakat:2006ki}. Note that the sequence termination condition is true only at a particular value $n$ since there is no guarantee that the sequence $\lambda_{n}$ is truly asymptotic.

\par Improvements to the efficiency of the AIM have been made which eliminates the need to take derivatives at each step in the iteration process, thereby reducing the computational overhead while also improving the overall accuracy of the method. This is done by reducing the AIM to a set of recursion relations which no longer involve differentiation. For more details on this improved AIM see Ref. \cite{Cho:2009wf}.


\section{Results and Concluding Remarks}\label{sec:5}

\begin{table}[t]
\caption{\textit{The QNM Frequencies for the Spin-$\frac{1}{2}$ Perturbation of the Kerr BH with rotation parameter $a = 0.00$, which is the Schwarzschild limit.}}\label{tab:1}
\begin{ruledtabular}
\begin{tabular}{|c|c|c|c|c|c|}
l & n & Numerical & Third Order WKB & Sixth Order WKB & AIM \\
\hline
0 & 0 & 0.1830 - 0.0970i & 0.1765 - 0.1001i & 0.1827 - 0.0949i & 0.1830 - 0.0969i \\
&   &   & (-3.55\% , 3.20\%) & (-0.16\% , -2.16\%) & ($<$0.01\% , -0.10\%) \\
\hline
1 & 0 & 0.3800 - 0.0964i & 0.3786 - 0.0965i & 0.3801 - 0.0964i & 0.3800 - 0.0964i \\
&   &   & (-0.37\% , 0.10\%) & (0.03\% , $<$0.01\%) & ($<$0.01\% , $<$0.01\%) \\
& 1 & 0.3558 - 0.2975i & 0.3536 - 0.2987i & 0.3559 - 0.2973i & 0.3568 - 0.2976i \\
&   &   & (-0.62\% , 0.40\%) & (0.03\% , -0.07\%) & (0.28\% , 0.03\%) \\
\hline
2 & 0 & 0.5741 - 0.0963i & 0.5737 - 0.0963i & 0.5741 - 0.0963i & 0.5741 - 0.0963i \\
&   &   & (-0.07\% , $<$0.01\%) & ($<$0.01\% , $<$0.01\%) & ($<$0.01\% , $<$0.01\%) \\
& 1 & 0.5570 - 0.2927i & 0.5562 - 0.2930i & 0.5570 - 0.2927i & 0.5573 - 0.2928i \\
&   &   & (-0.14\% , 0.10\%) & ($<$0.01\% , $<$0.01\%) & (0.05\% , 0.03\%) \\
& 2 & 0.5266 - 0.4997i & 0.5273 - 0.4972i & 0.5265 - 0.4997i & 0.5189 - 0.5213i \\
&   &   & (0.13\% , -0.50\%) & (-0.02\% , $<$0.01\%) & (-1.46\% , 4.32\%) \\
\hline
3 & 0 &  & 0.7672 - 0.0963i & 0.7674 - 0.0963i & 0.7674 - 0.0963i \\
&   &   &  & & ($<$0.01\% , $<$0.01\%) \\
& 1 &  & 0.7540 - 0.2910i & 0.7543 - 0.2910i & 0.7544 - 0.2910i \\
&   &   &  & & (0.01\% , $<$0.01\%) \\
& 2 &  & 0.7305 - 0.4909i & 0.7298 - 0.4919i & 0.7267 - 0.4928i \\
&   &   &  & & (-0.42\% , 0.18\%) \\
& 3 &  & 0.6999 - 0.6957i & 0.6967 - 0.7023i & \\
&   &   &  & & 
\end{tabular}
\end{ruledtabular}
\end{table}
\begin{table}[ht]
\caption{\textit{The QNM Frequencies for the Spin-$\frac{1}{2}$ Perturbation of the Kerr BH with a rotation parameter $a = 0.20$.}}\label{tab:2}
\begin{ruledtabular}
\begin{tabular}{|c|c|c|c|c|c|}
l & n & Numerical & Third Order WKB & Sixth Order WKB & AIM \\
\hline
0 & 0 & 0.1836 - 0.0967i & 0.1773 - 0.0997i & 0.1837 - 0.0942i & 0.1835 - 0.0965i \\
&   &   & (-3.43\% , 3.10\%) & (0.05\% , -2.59\%) & (-0.05\% , -0.21\%) \\
\hline
1 & 0 & 0.3811 - 0.0961i & 0.3796 - 0.0962i & 0.3810 - 0.0960i & 0.3809 - 0.0960i \\
&   &   & (-0.39\% , 0.10\%) & (-0.03\% , -0.10\%) & (-0.05\% , -0.10\%) \\
& 1 & 0.3572 - 0.2964i & 0.3550 - 0.2974i & 0.3572 - 0.2969i & 0.3583 - 0.2962i \\
&   &   & (-0.62\% , 0.34\%) & ($<$0.01\% , 0.17\%) & (0.31\% , -0.07\%) \\
\hline
2 & 0 & & 0.5750 - 0.0960i & 0.5754 - 0.0959i & 0.5754 - 0.0959i \\
&   &   & &  & ($<$0.01\% , $<$0.01\%) \\
& 1 & & 0.5578 - 0.2918i & 0.5586 - 0.2915i & 0.5589 - 0.2916i  \\
&   &   &  &  & (0.05\% , 0.03\%) \\
& 2 & & 0.5294 - 0.4951i & 0.5287 - 0.4975i &  \\
&   &   &  &  &   \\
\hline
3 & 0 &  & 0.7689 - 0.0959i & 0.7691 - 0.0950i & 0.7691 - 0.0959i \\
&   &   &  & & ($<$0.01\% , 0.95\%) \\
& 1 &  & 0.7559 - 0.2899i & 0.7563 - 0.2898i & 0.7563 - 0.2899i \\
&   &   &  & & ($<$0.01\% , 0.03\%) \\
& 2 &  & 0.7328 - 0.4889i & 0.7321 - 0.4899i & 0.7289 - 0.4910i \\
&   &   &  & & (-0.44\% , 0.22\%) \\
& 3 &  & 0.7028 - 0.6928i & 0.6997 - 0.6993i &  \\
&   &   &  & & 
\end{tabular}
\end{ruledtabular}
\end{table}
\begin{table}[ht]
\caption{\textit{The QNM Frequencies for the Spin-$\frac{1}{2}$ Perturbation of the Kerr BH with a rotation parameter $a = 0.40$.}}\label{tab:3}
\begin{ruledtabular}
\begin{tabular}{|c|c|c|c|c|c|}
l & n & Numerical & Third Order WKB & Sixth Order WKB & AIM \\
\hline
0 & 0 & 0.1854 - 0.0956i & 0.1798 - 0.0982i & 0.1844 - 0.0932i & 0.1851 - 0.0951i \\
&   &   & (-3.02\% , 2.72\%) & (-0.54\% , -2.51\%) & (-0.16\% , -0.52\%) \\
\hline
1 & 0 & 0.3843 - 0.0951i & 0.3825 - 0.0949i & 0.3838 - 0.0948i & 0.3837 - 0.0948i \\
&   &   & (-0.47\% , -0.21\%) & (-0.13\% , -0.32\%) & (-0.16\% , -0.32\%) \\
& 1 & 0.3614 - 0.2930i & 0.3592 - 0.2931i & 0.3612 - 0.2918i & 0.3630 - 0.2916i \\
&   &   & (-0.61\% , 0.03\%) & (-0.06\% , -0.41\%) & (0.44\% , -0.48\%) \\
\hline
2 & 0 &  & 0.5791 - 0.0948i & 0.5795 - 0.0947i & 0.5795 - 0.0947i \\
&   &   & &  & ($<$0.01\% , $<$0.01\%) \\
& 1 & & 0.5628 - 0.2880i & 0.5636 - 0.2878i & 0.5640 - 0.2877i  \\
&   &   &  &  & (0.07\% , 0.03\%) \\
& 2 & & 0.5357 - 0.4883i & 0.5352 - 0.4905i & 0.5472 - 0.5204i  \\
&   &   &  &  & (2.24\% , 6.10\%)  \\
\hline
3 & 0 &  & 0.7744 - 0.0947i & 0.7745 - 0.0947i & 0.7745 - 0.0947i \\
&   &   &  & & ($<$0.01\% , $<$0.01\%) \\
& 1 &  & 0.7620 - 0.2862i & 0.7624 - 0.2862i & 0.7625 - 0.2862i \\
&   &   &  & & (0.01\% , $<$0.01\%) \\
& 2 &  & 0.7400 - 0.4825i & 0.7395 - 0.4834i & 0.7358 - 0.4858i \\
&   &   &  & & (-0.50\% , 0.50\%) \\
& 3 &  & 0.7114 - 0.6835i & 0.7086 - 0.6893i &  \\
&   &   &  & & 
\end{tabular}
\end{ruledtabular}
\end{table}
\begin{table}[ht]
\caption{\textit{The QNM Frequencies for the Spin-$\frac{1}{2}$ Perturbation of the Kerr BH with a rotation parameter $a = 0.60$.}}\label{tab:4}
\begin{ruledtabular}
\begin{tabular}{|c|c|c|c|c|c|}
l & n & Numerical & Third Order WKB & Sixth Order WKB & AIM \\
\hline
0 & 0 & 0.1885 - 0.0934i & 0.1838 - 0.0953i & 0.1870 - 0.0909i & 0.1879 - 0.0925i \\
&   &   & (-2.49\% , 2.03\%) & (-0.80\% , -2.68\%) & (-0.32\% , -0.96\%) \\
\hline
1 & 0 & 0.3901 - 0.0931i & 0.3877 - 0.0925i & 0.3888 - 0.0924i & 0.3888 - 0.0924i \\
&   &   & (-0.62\% , -0.64\%) & (-0.33\% , -0.75\%) & (-0.33\% , -0.75\%) \\
& 1 & 0.3687 - 0.2861i & 0.3662 - 0.2849i & 0.3682 - 0.2837i & 0.3714 - 0.2811i \\
&   &   & (-0.68\% , -0.42\%) & (-0.13\% , -0.84\%) & (0.73\% , -1.75\%) \\
\hline
2 & 0 &  & 0.5866 - 0.0924i & 0.5869 - 0.0924i & 0.5869 - 0.0924i \\
&   &   & &  & ($<$0.01\% , $<$0.01\%) \\
& 1 & & 0.5715 - 0.2806i & 0.5724 - 0.2804i & 0.5730 - 0.2798i  \\
&   &   &  &  & (0.10\% , -0.21\%) \\
& 2 & & 0.5463 - 0.4751i & 0.5462 - 0.4768i &   \\
&   &   &  &  &  \\
\hline
3 & 0 &  & 0.7842 - 0.0924i & 0.7844 - 0.0924i & 0.7844 - 0.0924i \\
&   &   &  & & ($<$0.01\% , $<$0.01\%) \\
& 1 &  & 0.7728 - 0.2791i & 0.7732 - 0.2790i & 0.7734 - 0.2790i \\
&   &   &  & & (0.03\% , $<$0.01\%) \\
& 2 &  & 0.7524 - 0.4700i & 0.7522 - 0.4707i & 0.7493 - 0.4775i \\
&   &   &  & & (-0.39\% , 1.44\%) \\
& 3 &  & 0.7257 - 0.6652i & 0.7237 - 0.6700i &  \\
\end{tabular}
\end{ruledtabular}
\end{table}
\begin{table}[ht]
\caption{\textit{The QNM Frequencies for the Spin-$\frac{1}{2}$ Perturbation of the Kerr BH with a rotation parameter $a = 0.80$.}}\label{tab:5}
\begin{ruledtabular}
\begin{tabular}{|c|c|c|c|c|c|}
l & n & Numerical & Third Order WKB(J) & Sixth Order WKB(J) & AIM \\
\hline
0 & 0 & 0.1932 - 0.0891i & 0.1883 - 0.0896i & 0.1914 - 0.0865i & 0.1920 - 0.0872i \\
&   &   & (-2.54\% , 0.56\%) & (-0.93\% , -2.92\%) & (-0.62\% , -2.13\%) \\
\hline
1 & 0 & 0.3993 - 0.0893i & 0.3956 - 0.0881i & 0.3967 - 0.0880i & 0.3965 - 0.0880i \\
&   &   & (-0.93\% , -1.34\%) & (-0.65\% , -1.46\%) & (-0.70\% , -1.46\%) \\
& 1 & 0.3789 - 0.2728i & 0.3751 - 0.2701i & 0.3777 - 0.2687i &   \\
&   &   & (-1.00\% , -0.99\%) & (-0.32\% , -1.50\%) &   \\
\hline
2 & 0 &   & 0.5984 - 0.0881i & 0.5987 - 0.0881i & 0.5987 - 0.0882i \\
&   &   &   &   & ($<$0.01\% , 0.11\%) \\
& 1 &   & 0.5844 - 0.2669i & 0.5855 - 0.2667i & 0.5847 - 0.2644i \\
&   &   &   &   & (-0.14\% , -0.86\%) \\
& 2 &   & 0.5600 - 0.4512i & 0.5609 - 0.4517i &   \\
&   &   &   &   & \\
\hline
3 & 0 &  & 0.8000 - 0.0882i & 0.8001 - 0.0882i & 0.8001 - 0.0882i \\
&   &   &  & & ($<$0.01\% , $<$0.01\%) \\
& 1 &  & 0.7895 - 0.2659i & 0.7900 - 0.2658i & 0.7900 - 0.2655i \\
&   &   &  & & ($<$0.01\% , -0.11\%) \\
& 2 &  & 0.7702 - 0.4471i & 0.7706 - 0.4473i & \\
&   &   &  & & \\
& 3 &  & 0.7443 - 0.6320i & 0.7436 - 0.6345i &  \\
&   &   &  & & 
\end{tabular}
\end{ruledtabular}
\end{table}

\par Tables \ref{tab:1}, \ref{tab:2}, \ref{tab:3}, \ref{tab:4} and \ref{tab:5} constitute the main results of this paper, and contain the numerically computed QNM frequencies for varying values of the BH rotation parameter $a = \frac{J}{M}$, along with the semi-analytic WKB(J) (to third and sixth order) and AIM results. In each case, and where numerical results were available, they were compared with the semi-analytic solutions, these numerical results were taken from Refs. \cite{Jing:2005dt, Jing:2005pk}. At this point we re-iterate that we have set $c = G = 1$ and $M = 1$. Note that in the absence of numerical results, QNM frequencies calculated using the sixth order WKB(J) method were compared to the values from the AIM.

\par Unlike earlier cases, such as the case of scalar perturbations, previously published numerical values exist for the QNM frequencies that result from spin-half perturbations of a Kerr BH \cite{Jing:2005dt, Jing:2005pk}, at least for the angular numbers of $l = 0$ and $l = 1$. In comparing these numerical results, the accuracy of the WKB(J) method was worst (relative to the pure numerical evaluation) for parameter values $(a, l, n) = (0, 0, 0)$, with approximate errors of $(3.55\%, 3.20\%)$ at third order and $(0.16\%, 2.16\%)$ at sixth order, and showed steady improvement in accuracy with increasing $l$ and $n$, to $(0.13\%, 0.50\%)$ at third order and $(0.02\%, < 0.01\%)$ at sixth order. However, the AIM method experienced degraded accuracy with increasing $l$ and $n$ at $a = 0$, ranging from $(< 0.01\%, 0.10\%)$ at $l = n = 0$ to $(1.46\%, 4.32\%)$ at $l = n = 2$. However, the accuracy of the AIM was at its highest when $a$ was at its lowest, $n$ was at its lowest and $l$ was at its highest. Note that this for an iterative of only 15, though with this iterative depth the AIM was still significantly slower that the sixth order WKB(J) method.

\par For the QNM frequencies when $l = 2$ and $l = 3$, the AIM values were then compared to the sixth order WKB(J) method values. In these cases the accuracy of the AIM, as compared to the sixth order WKB(J) method was at its highest when $n$ was at its lowest and the $l$ values were higher, such as $(a, l, n) = (0, 3, 0)$, $(0.20, 2, 0)$, $(0.40, 2 \; or \; 3, 0)$, $(0.60, 2 \; or \; 3, 0)$ and $(0.80, 3, 0)$. Once again, numerical processors did evaluate significantly faster for the sixth order WKB(J) method than for the AIM.

\par In conclusion, we have, using the NP formalism, developed the Dirac equation in a Kerr background in a pedagogical manner. Using this result we have applied two semi-analytic techniques to calculate the QNM frequencies for comparison with available numerical results \cite{Jing:2005dt, Jing:2005pk} finding that the computational overhead for the WKB(J) method is lower at low $n$ than for the AIM. As such the WKB(J) method executes considerably more quickly than the AIM.


\acknowledgments

\par BJ would like to thank Roman Konoplya for his advice and useful discussions during the production of this work.


\end{document}